\documentclass[aps,pre,secnumarabic,nobalancelastpage,amsmath,amssymb,
nofootinbib,preprint]{revtex4-1}
\usepackage{graphics}      
\usepackage{graphicx}      
\usepackage{longtable}     
\usepackage{url}           
\usepackage{bm}            
\newcommand{\beq}{\begin{equation}}
\newcommand{\eeq}{\end{equation}}
\newcommand{\bea}{\begin{eqnarray}}
\newcommand{\eea}{\end{eqnarray}}

\begin{document}

\title{Thermal Casimir effect in Kerr spacetime with quintessence and massive gravitons}

\author{V. B. BEZERRA}\affiliation{Departamento de F\'isica, Universidade Federal da Para\'iba, CEP 58059-970, Caixa Postal 5.008, Jo\~ao Pessoa-PB, Brazil.}\email{valdir@fisica.ufpb.br}
\author{H. R. CHRISTIANSEN}\email{hugo.christiansen@ifce.edu.br}\affiliation{Departamento de F\'isica, Instituto Federal de Educa\c{c}\~ao, Ci\^encia e Tecnologia do Cear\'a (IFCE), Av. Dr. Guarany, 317, Derby Clube, CEP 62042-030, Sobral-CE, Brazil.}
\author{M. S. CUNHA}\email{marcony.cunha@uece.br}\affiliation{Grupo de F\'isica Te\'orica (GFT), Universidade Estadual do Cear\'a, Av. Dr. Silas Munguba, 1700, CEP 60714-903, Fortaleza-CE, Brazil.}
\author{C. R. MUNIZ}\email{celio.muniz@uece.br}
\author{M. O. TAHIM}\email{makarius.tahim@uece.br}\affiliation{Universidade Estadual do Cear\'a, Faculdade de Educa\c c\~ao, Ci\^encias e Letras do Sert\~ao Central, Planalto Universit\'ario, s/n, Quixad\'a-CE, Brazil.}
\begin{abstract}
Starting from an analytical expression for the Helmholtz free energy we calculate the thermal corrections to the Casimir energy-density and entropy within nearby ideal parallel plates in the vacuum of a massless scalar field. Our framework is the Kerr spacetime in the presence of quintessence and massive gravitons. The high and low temperature regimes are especially analysed in order to distinguish the main contributions. For instance, in the high temperature regime, we show that the force between the plates is repulsive and grows with both the quintessence and the massive gravitons. Regarding the Casimir entropy, our results are in agreement with the Nernst heat theorem and therefore confirm the third law of thermodynamics in the present scenario.

\end{abstract}

\maketitle

\section{Introduction}

The Casimir effect is, at its roots, the negative pressure acting between extremely close neutral ideal metallic plates in a perfect air void \cite{Casimir}. From the point of view of classical electrodynamics, there is no force acting between them; thus, the only possible explanation is that it results from the modifications of the quantum vacuum oscillations of the (zero-point) electromagnetic field determined by the plates boundaries. The effect is also found in strong gravitational field backgrounds in spacetimes with non-Minkowskian topology \cite{larry,DeWitt}. In such cases, there are in principle no material boundaries, but there exist some identification conditions imposed on the quantum fields which play the same role. Nowadays, we know that this phenomenon is not just inherent to the electromagnetic field but also to other vacua, such as the scalar, fermion, and tensor zero-point fields, and that it takes place between generic surfaces of real materials. See \cite{Trunov,Milton,Bordag} for a comprehensive review on the Casimir effect.
Among recent related work, we mention \cite{Elizalde1,Celio4,Quach,Stabile}.

Intensive research has been also devoted to the Casimir effect in cosmological models with non-trivial geometries and topologies (see e.g \cite{Bordag}). In this scenario, it is crucial to take into account the thermal corrections to the Casimir energy which will contribute to the energy-momentum tensor. This is mandatory for an appropriate description of the Universe at early stages, when the temperature was extremely high. Along this line of research, the total thermal stress-energy tensor of the scalar field was considered \cite{Klimchitskaya} in the Einstein cosmological model. The neutrino and electromagnetic fields were also considered in the Einstein and closed Friedmann cosmological models \cite{Mota}, in which cases , the total and Casimir free energy and the Casimir contributions to the stress-energy tensor were obtained, as well as their asymptotic behaviours at low and high temperatures.

For others works on this subject devoted to the study of thermal correction to the Casimir effect in the Einstein universe, we can mention \cite{Kleinert1,herondy1,herondy2}, as well as \cite{Kleinert2}, where more involved topologies were taken into account.

Likewise, in astrophysical scenarios, particularly those containing black holes and dark energy, it is important to calculate the thermal corrections to the Casimir energy and other thermodynamic quantities to understand the role played by the gravitational field, the boundaries, the space-time topology and temperature on the behaviour of the different quantum fields.

In the last two decades investigations report on the possible existence of a kind of cosmic \textit{dark} energy
which should be the principal substance of the Universe and the source of an accelerated global expansion. Although its origins and nature are still unknown its effect has been measured with accuracy \cite{PlanckSat}. There are theoretical models such as the $\Lambda$CDM, based on a cosmological constant, where dark energy has a constant density throughout the Universe. Others consider a space and time variable dark density modelled by physical fields dubbed \textit{quintessence} (see e.g \cite{Caldwell1,Doran} and references therein). At astrophysical scales, we can conjecture the existence of quintessential matter concentrated around stellar objects producing an additional gravitational shift of the light coming from distant stars \cite{Liu}. Such effects may be understood by means of particular analytical solutions of general relativity where quintessence matter surrounds static \cite{Kiselev} or rotating black holes \cite{Ghosh}, \cite{Toshmatov}. It has been shown that for an appropriate choice of some quintessential state parameters one can add other ingredients such as the source's electric charge. In the present paper, we will introduce quintessence together with
massive gravitons according to the proposal of de Rham, Gabadadze and Tolley (dRGT)  \cite{deRham:2010ik,deRham:2010kj}.

It is opportune to mention  that the inclusion of massive mediators of gravity into general relativity has been a difficult task for it is not just enough to introduce a term with the right momentum dimension obeying general covariance.
Since the first attempts, in 1939 \cite{Fierz:1939ix}, several corrections have been made along the years. Among these, significant progress took place in the early 1970s. Particularly, the discovery and treatment of the van Dam-Veltman-Zakharov discontinuity arising in the linear approximation \cite{VanNieuwenhuizen:1973fi,vanDam:1970vg,Zakharov:1970cc}
the origin of which is related to the validity of predictions inside the Vainshtein radius. Fortunately, the usual general relativity predictions can be recovered with a specific procedure of non-linear massive gravity \cite{Vainshtein:1972sx}. However, ghost instabilities were shortly found \cite{Boulware:1973my} for such non-linearity generates higher derivative terms in the equations of motion. It took thirty years to show that this issue could be circumvented by using the well known Stuckelberg mechanism \cite{ArkaniHamed:2002sp}. In order to avoid reappearance of ghosts in massive gravity, the set of allowed mass terms was restricted and built perturbatively in the dRGT model. The dRGT massive gravity is thus constructed so that the equations of motion have no higher derivative terms and the ghost field disappears. Despite these advances, the model is difficult to handle and analytical solutions can be hardly found.

Is gravity massive or not? If it is, which are the implications, local and cosmological? The central one is to change the behaviour of gravity at cosmological (huge) extents while keeping that of ordinary (massless) general relativity at shorter distances. In a sense, this can be faced as a bimetric model which may answer the dilemma of accelerated expansion of the Universe, viz. the dark energy or quintessence problem. Its connection with the Casimir effect can certainly shed some light on this puzzle (see for instance \cite{Leandros} where tit has been considered to deal with the cosmological constant problem).

With this idea in mind, here we calculate the Casimir energy at finite temperature in the quantum vacuum of a scalar field in a cavity orbiting a rotating spherical body surrounded by quintessence. Furthermore, we will assume the presence of gravitons with non-zero mass in such a scenario. This setup is in order to investigate the effects of both the relativistic gravitational attraction and dragging of spacetime around the body, as well as the implications of quintessence and massive gravitons on the regularized vacuum energy of a massless scalar field. Our study follows that in \cite{Sorge2} which analysed the gravitational and rotational effects due to Kerr spacetime on that energy. At the same time, we generalize very recent outcomes \cite{Celio5} where thermal effects are considered. In Section II we present general considerations on the Casimir effect in the Kerr black hole in the presence of quintessence and massive gravitons.  Section III is devoted to discuss the thermal corrections to the Casimir effect and present the results concerning Casimir energy and entropy. In Section IV we present our conclusions.

\section{Casimir effect in the Kerr black hole surrounded by \\ quintessence and massive gravitons}

The spacetime generated by an axisymmetric gravitational source of mass $M$ and angular momentum $J$, surrounded by quintessence, can be described by an extension of a metric recently found by Ghosh \cite{Ghosh} obtained by means of Newman-Janis complex transformations made on the static counterpart of a black hole with quintessence as solved by Kiselev \cite{Kiselev}. Here, the quintessence and the graviton mass are characterized by a set of state parameters $\varpi_i$ and $\overline{\alpha}_i$. In Boyer-Linquidist coordinates, the extended metric results
\begin{equation}
ds^2\!=\!\left(\!\frac{\widetilde{\Delta}-a^2\sin^2{\theta}}{\Sigma}\!\right)\!\!dt^2\!+\!
2a\!\left(\!1-\frac{\widetilde{\Delta}-a^2\sin^2{\theta}}{\Sigma}\!\right)\sin^2{\theta}
dtd\phi-\frac{\Sigma}{\widetilde{\Delta}} dr^2-\Sigma d\theta^2-\frac{\widetilde{A}}
{\Sigma}\sin^2{\theta} d\phi^2, \label{KerrQuintMetr}
\end{equation}
where
\begin{eqnarray}
&a=J/M, \,\,\,\,\,\ \Sigma=r^2+a^2\cos^2\!\theta, & \nonumber \\
&\widetilde{\Delta}=r^2+a^2-2Ma\,r-\sum_{i={o,1,2,3}}\overline{\alpha}_i\Sigma^{(1-3\varpi_{i})/2}, &
\\  \label{KerrParam}
&\widetilde{A}=\Sigma^2+a^2\sin^2{\theta}\left(2\Sigma-\widetilde{\Delta}+a^2\sin^2{\theta}\right)\!,
& \nonumber
\end{eqnarray}
and we use units such that $c= \hbar= G=1$.
The first of the new parameters, $\varpi_o$,  describes the
state of inflational quintessence which, considering the actual accelerated
expansion of the Universe pushed by dark energy (de Sitter outer horizon)
obeys $\varpi_o\in(-1,-1/3)$. The interval $\varpi_o\in(-1/3,0)$ corresponds to a flat inner horizon; the special values $\varpi_o=1/3$, $\varpi_o=0$, and $\varpi_o=-1$ correspond respectively to a universe with dominance of relativistic matter, dust matter, or a cosmological constant; the values
$\varpi_o<-1$ lead to a so-called phantom energy \cite{Caldwell}.

The parameter $\overline{\alpha}_o$ is such that the state equation for the quintessential
mass density is given by $\rho_q=\varpi_o p_q$ where
\begin{equation}\label{quintpress}
p_q=-\frac{3\overline{\alpha}_o}{2r^{3(\varpi_o+1)}}.
\end{equation}
is the quintessence pressure \cite{Kiselev}.
The other state parameters in Eq.(2), $\varpi_1,\varpi_2,\varpi_3$, describe entities
identified with elements of the dRGT model of massive gravity in which the gravitons
have mass $m_g$ \cite{Ghosh} which we associate with the corresponding $\overline\alpha_i$.
Thus, $\varpi_{1}=-1/3$ and $\overline\alpha_1=\zeta m^2_g$ are identified with a global
monopole-like potential term which appears as a constant correcting the Newtonian one;
$\varpi_{2}=-2/3$ and $\overline\alpha_2=\gamma m^2_g$ are associated with a potential
term linear in $r$; and $\varpi_{3}=-1$, $\overline\alpha_3=\lambda m^2_g$ are linked
to a cosmological constant-like term \cite{Ghosh1}, which overlap with $\varpi_o=-1$
state parameter of the quintessence itself. The parameters $\zeta$, $\gamma$ and
$\lambda$ are proportionality constants, with units of (length)$^2$, (length) and
dimensionless, respectively. Later in this work we will set $\lambda=\zeta=0$ in
order to infer bounds on the graviton mass from the Casimir experiments on Earth.
It is worth to notice that for  $\overline{\alpha}_{o,1,2,3}=0$, Eq. (\ref{KerrQuintMetr})
reduces to the standard Kerr black hole where $\widetilde{\Delta}=\Delta$ and $\widetilde{A}=A$.
In the particular case $a=0$, the metric yields the Schwarzschild solution surrounded
by several kinds of quintessential matter according to the Kiselev description \cite{Kiselev}.

Here, we also extend the work of Sorge \cite{Sorge2} which
studied the vacuum energy of a scalar massless field confined in a Casimir cavity moving in a
circular equatorial orbit in the exact Kerr space-time geometry.
The cavity is in a locally co-moving referential frame where
Cartesian coordinates $(x, y, z)$ are centred on one of the plates
so that the $z$ axis is
tangential to the path of the circular orbit. 
Thus, the relation between the spherical coordinates centred in the source
and the Cartesian axes of the orbiting system  with angular velocity $\Omega$,
is $ dy = dr$, $ dz = r\,d\phi' $ and $ dx = -r\,d\theta$, where $\phi'=\phi-\Omega t$.
Therefore, in the orbiting Cartesian frame the metric becomes
\begin{equation}
ds^2 =\widetilde{C}^{-2}\left(\Omega\right)\!dt^2-\frac{\widetilde{A}}{r \Sigma}\sin^2\!\theta\left(\widetilde{\omega}_{d}-\Omega\right) dtdz-
\frac{\Sigma}{\widetilde{\Delta}} dy^2-\frac{\Sigma}{r^{2}} dx^2-
\frac{\widetilde{A}}{r^{2}\Sigma}\sin^2\!\theta dz^2, \label{01}
\end{equation}
where $\widetilde{\omega}_{d}$ is the angular velocity of the dragging
of the spacetime around the gravitational source
\begin{equation}
\widetilde{\omega}_{d}=-\frac{g_{t\phi}}{g_{\phi\phi}}\,=\frac{2Mar}{\widetilde{A}},\label{02}
\end{equation}
and
\begin{equation}
\widetilde{C}^{-2}\left(\Omega\right)=\frac{\Sigma \widetilde{\Delta}}{\widetilde{A}}\left[1-\frac{\widetilde{A}^{2}}{\widetilde{\Delta} \Sigma^2}
\sin^2{\theta}\left(\Omega-\widetilde{\omega}_d\right)^2\right]. \label{03}
\end{equation}
With these equations, we can calculate the normal modes, $\omega_{n}$, and the Casimir energy.
Note that considering $\overline{\alpha}=0$ (viz. no quintessence) removes
the tilde in the above expressions, and further specializing  $\theta=\pi/2$
retrieves the results in \cite{Sorge2}.

The massless scalar field obeys the covariant Klein-Gordon equation, given by
\begin{equation}
\left[\frac{1}{\sqrt{-\hat{g}}}\partial_{\mu}
\left(\sqrt{-\hat{g}}\hat{g}^{\mu \nu}\partial_{\nu}\right)+\xi R\right]
\phi\left(t,\textbf{x} \right)=0, \label{04}
\end{equation}
where $\hat{g}^{\mu\nu}$ is the metric given by the inverse matrix of Eq. (\ref{01})
which can be considered almost constant inside the cavity since $L \ll\, r$.
Assuming minimal coupling  ($\xi = 0$) for simplicity,  it leads to
\begin{equation}
\hat{g}^{\mu\nu}\,\partial_{\mu}\partial_{\nu}\,\phi(t,\textbf{r})=0.
\end{equation}

The eigenfunctions of the scalar field confined to a cavity limited
by parallel plates orbiting the black hole, $\phi_n$, read \cite{Celio5}
\begin{equation} \label{05}
\phi_{n}\left(t,x,y,z\right)=N_{n}e^{i\left(k_{x}x+k_{y}y-\omega_n t\right)}
e^{i\beta_{n}z}\sin{\!\left(\frac{n \pi}{L}z\right)},
\end{equation}
where
\begin{equation}\label{05.a}
\beta_{n}=\frac{\widetilde{A}\left(\Omega-\widetilde{\omega}_{d}\right)\sin^{2}{\theta} \widetilde{C}^{2}_{\Omega}}{r\Sigma}\,\omega_n,
\end{equation}
and $N_{n}$ is the normalization constant, given by
\begin{equation}
N_{n}^{2}=\frac{r^3}{\left(2\pi\right)^{2}L\omega_{n}\Sigma^2\sin{\theta}}
\sqrt{\frac{\widetilde{A}}{\widetilde{\Delta}}}\,\widetilde{C}^{-3}_{\Omega}. \label{06}
\end{equation}
The eigenfrequencies of the confined field are
\begin{equation}
\omega_{n}=\frac{r}{\sqrt{\widetilde{\Delta}}\sin\!\left(\theta\right)  \widetilde{C}^{2}_{\Omega}}\left[\left(\frac{n\pi}{L}\right)^{2}+
\frac{\widetilde{\Delta}\sin^2{\theta}  \widetilde{C}^{2}_{\Omega}}{\Sigma}\left(k_{x}^{2}+
\frac{\widetilde{\Delta}}{r^{2}}k_{y}^{2}\right)\right]^{\frac{1}{2}}. \label{07}
\end{equation}

Since the proper length of the cavity is
\beq L_p=\widetilde{C}_{\Omega}\frac{\sin{\theta}\,\sqrt{\widetilde{\Delta}}}{r}\,L,\eeq
the regularized Casimir energy density in a cavity orbiting a massive spherical
rotating body surrounded by quintessence reads
\begin{equation}
\langle\epsilon_{vac}\rangle^{(ren)}\!=-\frac{\pi^2}{1440L_{p}^{4}}
\sqrt{\frac{\Sigma}{r^2}}\left[1-\frac{\widetilde{A}^{2}}{\widetilde{\Delta} \Sigma ^2}\sin^2{\theta}\left(\Omega-\widetilde{\omega}_d\right)^2\right]^{\frac{1}{2}}. \label{08}
 \end{equation}
It is easy to verify that when the cavity is at the north pole the Casimir energy density is simply
  \begin{equation}
\langle\epsilon_{vac}\rangle^{(ren)}\!=-\frac{\pi^2}{1440L_{p}^{4}}\sqrt{\frac{r^2+a^2}{r^2}}, \label{09}
 \end{equation}
which depends just on the angular momentum of the source and have no traces of the quintessential matter. In the equatorial plane ($\theta=\frac{\pi}{2}$) the result found in \cite{Sorge2} is recovered for $\overline{\alpha}=0$. It is worth noting that turning off the rotation of the source ($a=0$) retrieves the Casimir energy of the massless scalar field in the Minkowski space vacuum. Notice also that the expressions (\ref{08}) and (\ref{09}) also correct Eq.(34) and Eq.(35) in Ref.\cite{Celio5}
which had both a (wrong) inversion in the rooted factor.

\section{Thermal corrections to the Casimir energy}
The renormalized Helmholtz free energy associated with the vacuum in a finite volume $V$ is given by \cite{Bordag}
\begin{equation}
\tilde{F}_0 = E_{0}^{(ren)}+\Delta_T F_0 -\mathcal{O}(T^2)-\mathcal{O}(T^3)- V\,f_{bb}, \label{Eq.51}
\end{equation}
where $E_{0}^{(ren)}=V_{p}\,\langle\epsilon_{vac}\rangle^{(ren)}$
is the temperature-independent renormalized vacuum energy, $V_p$ being the proper volume of the cavity.
The other terms are
\begin{equation}
\Delta_T F_0 =  S_0k_B T\int \frac{d^2{\bf k}}{(2\pi)^2}\sum_{j}^{\infty}\ln\left(1-e^{-\frac{\omega_j}{k_B T}} \right)\ \ \ \mbox{and} \ \ \ f_{bb}=k_B T\int \frac{d^3\mathbf{k}}{(2\pi)^3}\ln\left(1-e^{-\frac{\omega(k)}{k_B T}} \right),\label{Eq.52}
\end{equation}
where $S_0$ is the area of the plates, $k_B$ is the Boltzmann constant, $\omega_j$ are the frequencies of the normal modes $\phi_j\left(\mathbf{r},t\right)$ and $k$ is the modulus of the wave vector $\mathbf{k}$. The term $Vf_{bb}$ is the usual black body free energy, and $\mathcal{O}(T^2)$, $\mathcal{O}(T^3)$ are obtained from expansion of the free energy for high temperatures. These terms, associated with the black body free energy of lines, surfaces and volumes, are subtracted from $\Delta_T F_0$ in order to renormalize the thermal vacuum energy \cite{Bordag}. In fact, we will see that there is no $\mathcal{O}(T^2)$ term.

Our aim here is obtaining the thermal energy as well as other thermodynamic variables, as pressure and entropy,
associated with the vacuum between the plates, which has proper volume $V_p=S_p\,L_p$, with $S_p=\int\int\sqrt{g_{xx}g_{yy}}dxdy=(\Sigma/\sqrt{\widetilde{\Delta}}r)S_0$.

Replacing the normal frequencies given by Eq. (\ref{07}) into the first of Eqs. (\ref{Eq.52}) we get
\begin{eqnarray}
\Delta_T F_0 &=& \frac{S_0 k_B T}{(2\pi)^2}\sum_{n=0}^{\infty}\int_{0}^{\infty}\int_{0}^{\infty}dk_xdk_y\\\nonumber
&\times&\ln\left\{1-\exp{\!\left[\frac{-\widetilde{C}^{-1}\left(\Omega,\theta\right)}{k_BT}
\sqrt{\frac{r^{2}}{\sin^2{\theta}\widetilde{\Delta} \widetilde{C}^{2}\!
\left(\Omega,\theta\right)}\left(\frac{n\pi}{L}\right)^{2}+
\frac{r^2}{\Sigma}k_{x}^{2}+\frac{\widetilde{\Delta}}{\Sigma}k_{y}^{2}}\,\right]} \right\},\label{Eq.56}
\end{eqnarray}
since the wave vector has two continuous components parallel to the plates $(k_x,k_y)$ and one discrete perpendicular to them  $\left(k_z \right)$.

Making the transformations $\frac{r}{\sqrt{\Sigma}}k_{x}=\widetilde{k}_{x}$, $\sqrt{\frac{\widetilde{\Delta}}{\Sigma}}k_{y}=\widetilde{k}_{y}$, and expanding the logarithm one obtains
\begin{equation}
\Delta_T F_0 =- \frac{S_0\Sigma k_B T }{2\pi r\sqrt{\widetilde{\Delta}}}\sum_{n=0}^{\infty}\sum_{s=1}^{\infty}\frac{1}{s}\int_{0}^{\infty}
\widetilde{k}_{||}\, e^{-\frac{s\widetilde{C}^{-1}\!\left(\Omega,\theta\right)}{k_B T}
\left[\frac{r^{2}}{\widetilde{\Delta} \sin^2{\theta}\widetilde{C}^{2}\!
\left(\Omega,\theta\right)}\left(\frac{n\pi}{L}\right)^{2}+
\tilde{k}_{||}^2\right]^{\frac{1}{2}}}d\widetilde{k}_{||}, \label{Eq.58}
\end{equation}
where, on the parallel plates, $2\pi\widetilde{k}_{||}d\widetilde{k}_{||}=
d\widetilde{k_x}d\widetilde{k_y}$.
Changing the integration variable in the form $\frac{r^{2}}{\widetilde{\Delta} \sin^2{\theta}\widetilde{C}^{2}\!\left(\Omega,\theta\right)}\left(\frac{n\pi}{L}\right)^2+
k_{\parallel}^2=\frac{r^{2}}{\widetilde{\Delta} \sin^2{\theta}\widetilde{C}^{2}\!\left(\Omega,\theta\right)}\left(\frac{n\pi}{L}\right)^2 u^2$, we get
\begin{equation}
\Delta_T F_0 =-\frac{S_0 k_B T}{2 L^2}\frac{\pi\Sigma r}{\sin^2{\theta}\sqrt{\widetilde{\Delta}^3} \widetilde{C}^{2}\!\left(\Omega,\theta\right)}\sum_{n=0}^{\infty}\sum_{s=1}^{\infty}\frac{n^2}{s}
\int_{1}^{\infty}e^{-\frac{s n \pi r \widetilde{C}^{-2}\!
\left(\Omega,\theta\right)}{k_B T\sin{\theta}\sqrt{\widetilde{\Delta}} L}u}udu. \label{Eq.60}
\end{equation}
The free energy density of the black body is given by the second of Eqs. (\ref{Eq.52}),
and can be directly solved:
\begin{equation}
f_{bb}=-\frac{\sin{\theta}\Sigma\widetilde{C}^4(\Omega,\theta)}{r^2}\frac{\pi^2(k_BT)^4}{90}.\label{Eq.62}
\end{equation}	
Thus, the total free energy (\ref{Eq.51}) is
\begin{eqnarray}
\tilde{F}_0\!=\! E_{0}^{(ren)} &-&V_p\sum_{n=0}^{\infty}\sum_{s=1}^{\infty}\left[\frac{n}{s^2}
\frac{\widetilde{C}(\Omega,\theta)(k_BT)^2}{2L_p^2}+\frac{1}{s^3}\frac{\widetilde{C}^2(\Omega,
\theta)(k_BT)^3}{2\pi L_p}\right]\exp{\left(\frac{-sn\pi}{\widetilde{C}(\Omega,\theta)L_pk_BT}\right)}\nonumber\\
&-&\mathcal{O}(T^2)-\mathcal{O}(T^3)+V_p\frac{\widetilde{C}^3(\Omega,\theta)\pi^2(k_BT)^4}{90}, \label{Eq.64}
\end{eqnarray}
where the integral in Eq. (\ref{Eq.60}) has been exactly solved and we have used
 its value for both the proper length and plates area.

\subsection{Thermal Casimir energy}

Defining the free energy density,  $\tilde{f}_0=\frac{\tilde{F}_0}{V_p}$,
and considering that the density of Casimir energy with thermal corrections is given by
\begin{equation}
u_{0}^{(ren)}=\langle\epsilon_{vac}\rangle^{(ren)}\!-T^2\frac{\partial}{\partial T}
\left(\frac{\tilde{f}_0}{T}\right), \label{Eq.65}
\end{equation}
with the aid of Eq. (\ref{08}) we arrive at
\begin{eqnarray}
u_{0}^{(ren)} &=&-\frac{\pi^2}{1440L_{p}^{4}}\sqrt{\frac{\Sigma}{r^2}}
\left[1-\frac{\widetilde{A}^{2}}{\widetilde{\Delta} \Sigma ^2}\sin^2{\theta}\left(\Omega-\widetilde{\omega}_d\right)^2\right]^{\frac{1}{2}}-
\frac{\widetilde{C}^3(\Omega,\theta)\pi^2(k_BT)^4}{30}\nonumber \\ & +&\sum_{n,s=1}^{\infty}
\left[\!\frac{n^2\pi (k_{B}T) }{2sL_{p}^3}+\frac{n\widetilde{C}(\Omega,\theta)\left(k_{B}T
\right)^2}{s^2L_{p}^2}+
\frac{\widetilde{C}^2(\Omega,\theta)\left(k_{B}T\right)^{3}}{s^3 \pi L_p}\right]
e^{-\frac{s n \pi}{k_{B}T\widetilde{C}\!\left(\Omega,\theta\right)\!L_{p}}},\nonumber\\ \label{Eq.67}
\end{eqnarray}
minus a renormalization term proportional to $T^3$ about which we will comment below.
Notice that the {thermal} correction to the Casimir energy
 depends on the quintessence as well as on the graviton
mass even at the north pole, in contrast to the $T=0$ case where both disappear.

Looking at Eq. (\ref{Eq.67}), we see that the terms
dominating $u_{0}^{(ren)}$ at high temperatures
are a $T^3/L_p$ term and the black body subtraction.
The former is not only of lower degree
but is also irrelevant for the thermal Casimir energy,
obtained by multiplying the energy density by the proper
volume, will no longer depend on $L_p$.
Thus, the purely quantum thermal Casimir energy is given by
\begin{equation}
U_{0}^{(ren)}(T) \approx-S_pL_p\frac{\widetilde{C}^3(\Omega,\theta)\pi^2(k_BT)^4}{30}.
\label{CasimirEnergyHT}
\end{equation}
The black body contribution to the regularized vacuum energy therefore
 corresponds to a constant repulsive force between the plates.

Applying this result to a system with weak gravity and
weak effects of rotation ($a^2\rightarrow 0$), considering also
$\Omega=\tilde{\omega}_d$,
we obtain from Eqs.(\ref{KerrParam}) and (\ref{03})
\begin{equation}\label{C3}
\tilde{C}^3\approx 1+3M/r+(r/R_q)^{-(1+3\varpi_o)}+(3/2)\gamma m_g^2\,r,
\end{equation}
where we have assumed $\gamma \geqslant 0$ and made $\lambda=\zeta=0$ for simplicity \cite{Ghosh1}.
Here, $R_q$ is a characteristic length or scale in
which the quintessence yields measurable effects \cite{Kiselev}. Thus, equations (\ref{CasimirEnergyHT}) and (\ref{C3}) show that at high temperatures the Casimir effect is destroyed. In fact, the positive value of the Casimir pressure,
\begin{equation}\label{pressure}
 -\partial U_{0}^{(ren)}(T)/\partial V_p\thickapprox\frac{\widetilde{C}^3(\Omega,\theta)\pi^2(k_BT)^4}{30},
\end{equation}
indicates a repulsion between the plates that
gets even larger with the presence of quintessence. From the details of Eq. (\ref{C3})
we see that the massive gravity term also gives rise to repulsive forces at high temperatures,
instead of expected.

\begin{figure}[!h]
\centering
\includegraphics[scale=0.45]{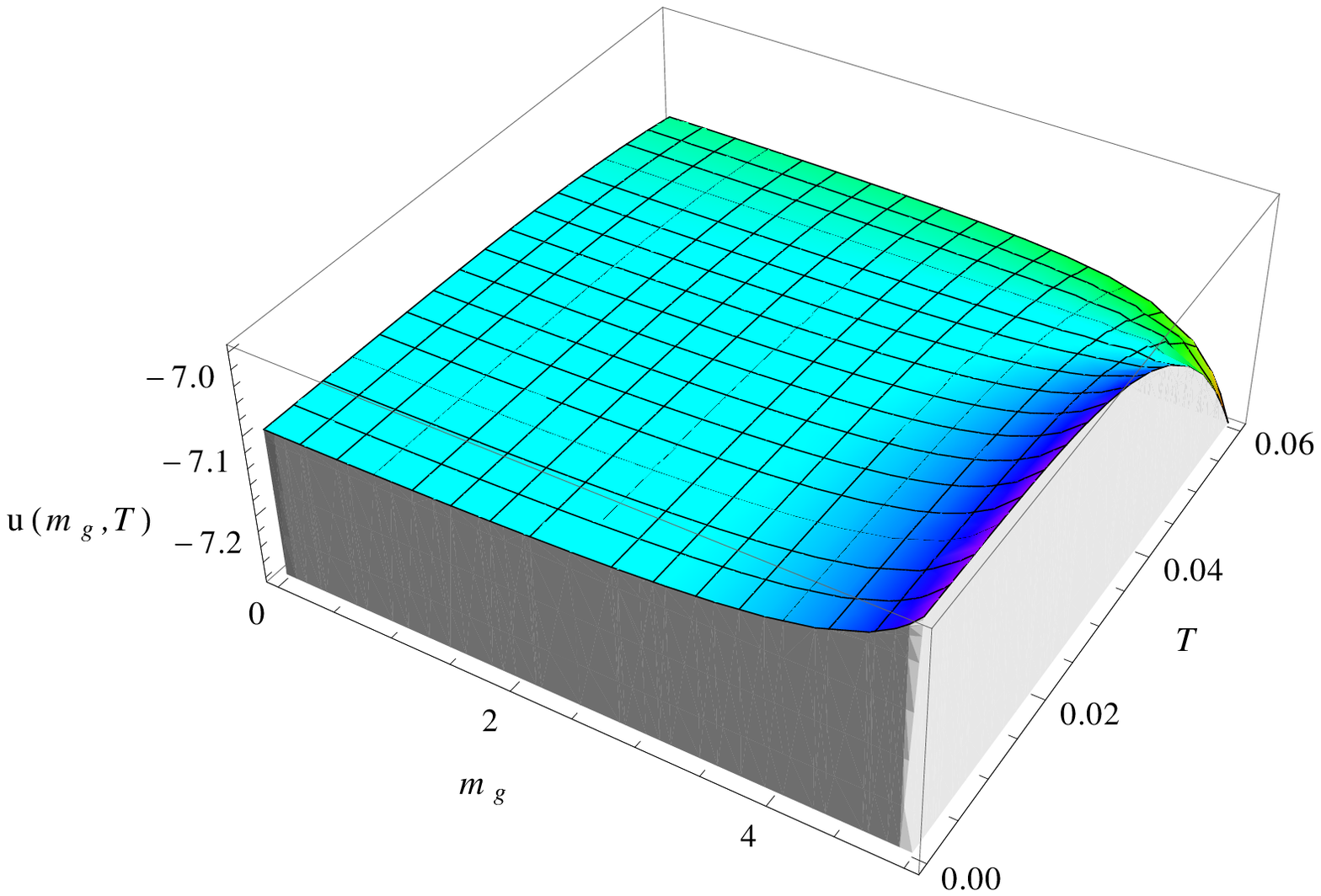}
\includegraphics[scale=0.4]{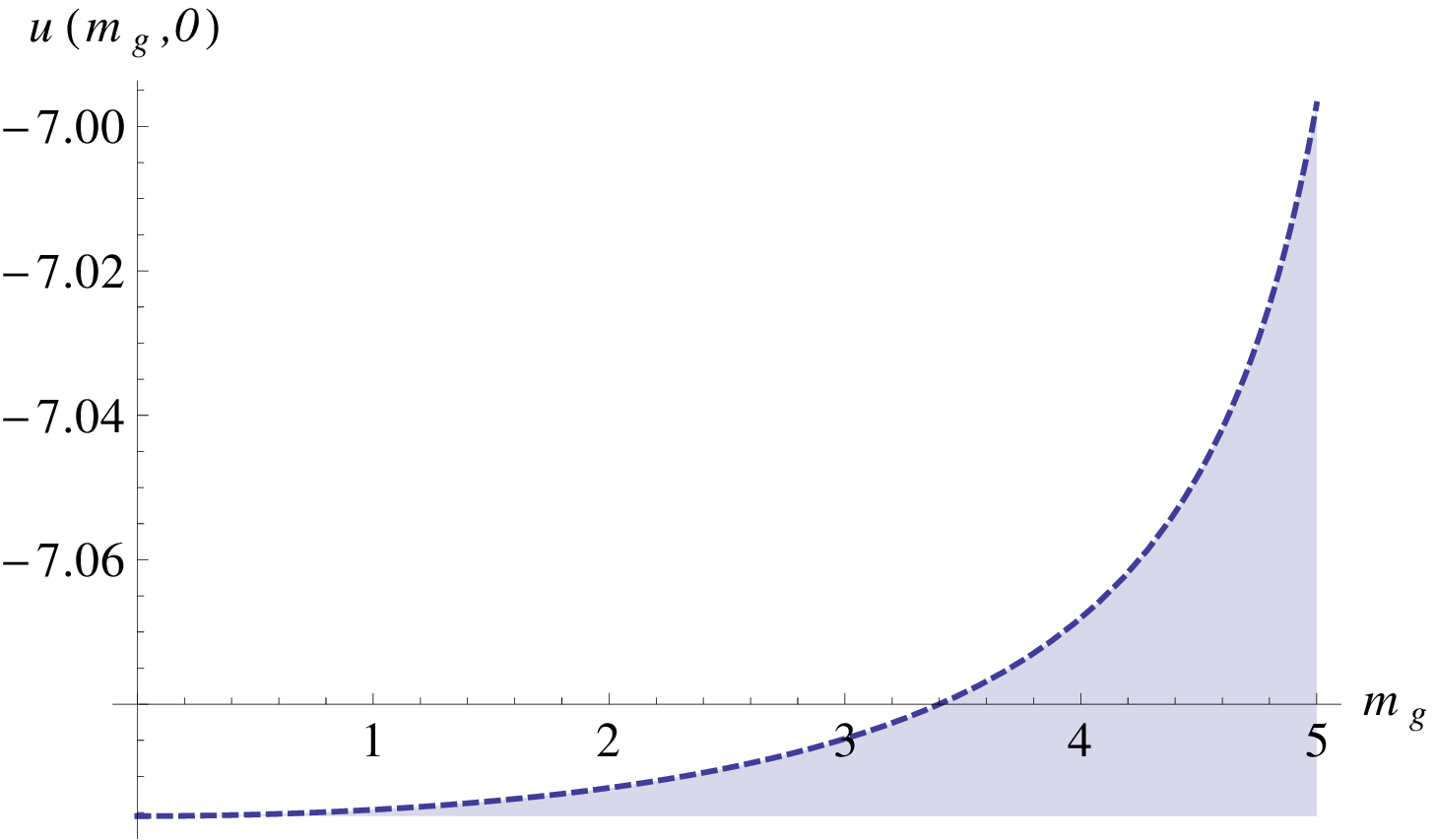}
\caption{$u_0(m_g,T)$ and $u_0(m_g,0)$ depicted at the equator (multiplied by $10^{3}$)  for a set of parameters: $r=10$, $\Omega=2\omega_d$, $\theta=\pi/2$, $\alpha_o=0.001$, $\varpi_o=-1/3$, $\varpi_2=-2/3$, $\alpha_2=\gamma m_g^2$, $\gamma=10^{-3}$, $L_p=k_B=1$.} \label{umg}
\end{figure}

\begin{figure}[!h]
\centering
\includegraphics[scale=0.4]{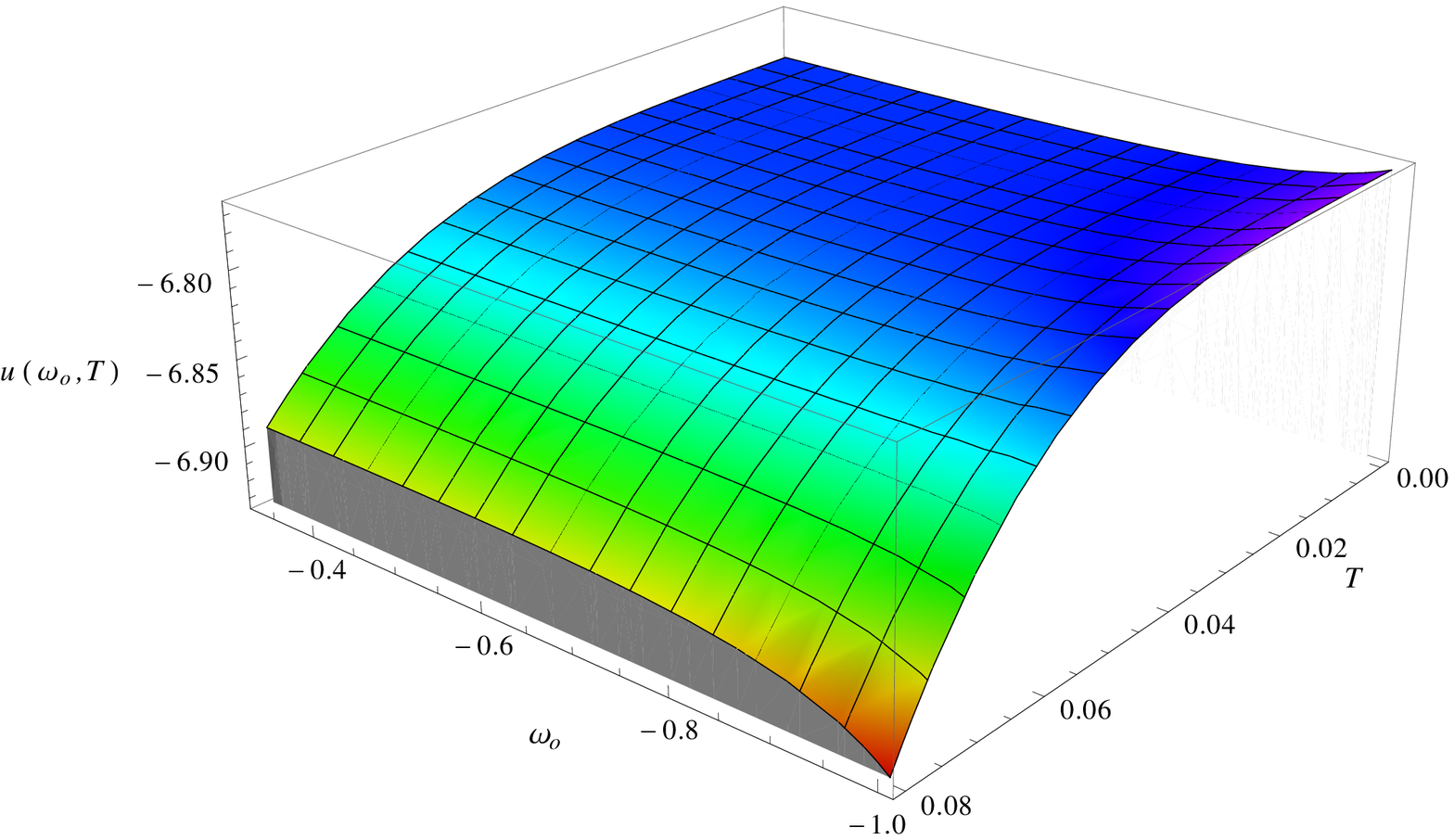}
\includegraphics[scale=0.4]{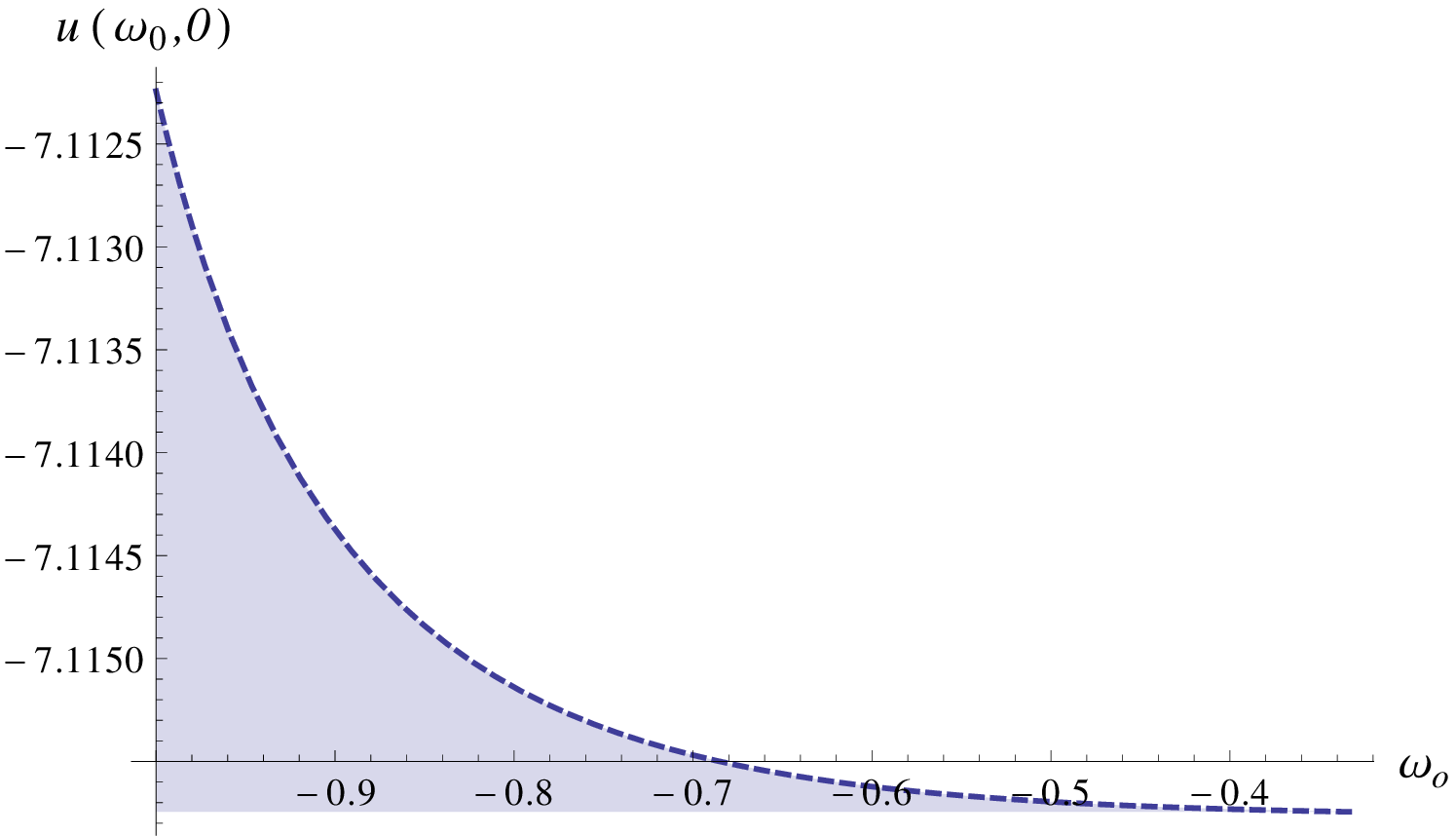}
\caption{$u_0(\varpi_o,T)$ and $u_0(\varpi_0,0)$ depicted at the equator (multiplied by $10^{3}$)  for a set of parameters: $r=10$, $\Omega=2\omega_d$, $\theta=\pi/2$, $\alpha_o=0.001$, $\varpi_2=-2/3$ $, \alpha_2=\gamma m_g^2$, $\gamma=10^{-3}$, $m_g=1$, $L_p=k_B=1$.} \label{uwo}
\end{figure}

By taking into account the low temperature regime, Eq. (\ref{Eq.67})
yields a Casimir energy density given by
\begin{eqnarray}
u_{0}^{(ren)} &\approx&-\frac{\pi^2}{1440L_{p}^{4}}\sqrt{\frac{\Sigma}{r^2}}
\left[1-\frac{\widetilde{A}^{2}}{\widetilde{\Delta}
\Sigma^2}\sin^2{\theta}\left(\Omega-\widetilde{\omega}_d\right)^2\right]^{\frac{1}{2}}
-\frac{\widetilde{C}^3(\Omega,\theta)\pi^2(k_BT)^4}{30} \nonumber\\
&+&
\left(\!\frac{\pi k_{B}T }{2L_{p}^3}\right)e^{-\frac{\pi}{k_{B}T\widetilde{C}\!\left(\Omega,\theta\right)\!L_{p}}} +\dots
 \label{Eq.69}
\end{eqnarray}
For $T\simeq 0$, the force resulting from the first term is in principle attractive, as expected, but
the second is repulsive. The following terms have both kinds of components
but are heavily smashed by the exponential at low $T$. 
In fact, the resulting sign of the force depends non-trivially
on several parameters. Therefore, at low temperature quintessential matter and graviton mass
can intensify or attenuate the force between the plates
and it can be either attractive or repulsive depending also on the distance,
the temperature and the rotational parameters, see Fig.\ref{umg} and \ref{uwo}.


\subsection{Casimir Entropy}

The Casimir entropy, $\mathcal{S}_{(C)}=-\frac{\partial \tilde{F}_0}{\partial T}$,
can be computed from the Helmholtz free energy given by Eq. (\ref{Eq.64}).
In the high temperature approximation it is given by
\begin{equation}\label{entropyHT}
\mathcal{S}_{(C)}\thickapprox
-V_p\frac{2\widetilde{C}^3(\Omega,\theta)\pi^2k_B^4T^3}{45}.
\end{equation}
It is worth pointing out that such negative contribution here arises from the fact
that we are dealing with an open system (namely,  just part of a closed system).

We see that the presence of quintessential matter, as included in $\widetilde{C}^3$,
contributes to diminish the entropy of the system,
while the massive gravitons contribute to increase the entropy.

At low temperatures, we find that the leading terms are
\begin{equation}
\mathcal{S}_{(C)}\approx S_0\,\zeta(0)\zeta(3)\,\frac{\widetilde{C}^2\,k_B^3T^2}{2\pi}+
2S_0\,\widetilde{C}\,\frac{k_B^2T}{2L_p}\,e^{-\frac{\pi}{\widetilde{C}L_pk_BT}},
\end{equation}
where $\zeta(0)=-1/2$ and $\zeta(3)\approx 1.202$.
From the expression above, we conclude that the Casimir
entropy is always growing as expected, and
we verify that $\lim_{T\rightarrow 0} \mathcal{S}_{(C)} = 0$ {\it i.e.}
the third law of thermodynamics is satisfied.

\section{Concluding Remarks}

We have obtained the thermal corrections to the Casimir energy of a massless scalar vacuum within parallel plates
in a Kerr spacetime filled with quintessence and dRGT massive gravitons. This generalization
has been performed following the procedure
developed in the papers of Kiselev \cite{Kiselev} and Ghosh et al \cite{Ghosh1}.
Our work extends Ref. \cite{Celio5} where quintessence was considered in massless gravity at zero temperature. We have also extended the results of Zhang \cite{Zhang} in which thermal corrections were computed although, conversely,
in the absence of both quintessence or massive gravitons.
Furthermore, we have obtained expressions which hold for arbitrary angles,
not just restricted to the equatorial plane as in recent attempts \cite{Sorge2,Zhang}.

Indeed, we have shown that both massive gravitons and quintessence
contribute to the regularized vacuum energy, even when the plates
are at the north pole and this is different from what happens at zero temperature.
At very high temperatures we have shown that the cavity effectively behaves as a black body radiator
with a repulsive force between the plates. Thus, by considering a gravitational source endowed with weak field and low rotation, we have also inferred that such a repulsive force grows with both the presence of the quintessential matter and the massive gravitons.

At low temperatures, we have distinguished two main contributions to the pressure in the cavity.
One is independent of the temperature and, as expected, produce an attractive force.
We observe an inverse fourth-order  dependence with the distance, $L_p$, and a more involved
dependence with quintessence and the gravitonic mass.
The second term is the above mentioned $T^4$ repulsive contribution which although
less significant in this regime is a new feature to be considered. The dependence of the Casimir force
with the mass of the graviton and the quintessence matter is in fact quite involved. It can
be appreciated in detail in the figures as a function of  temperature. We see that even the sign of the force
can flip for different values of the parameters.

\section*{Acknowledgements}

The authors would like to thank Conselho Nacional de Desenvolvimento Cient\'{i}fico e Tecnol\'{o}gico (CNPq) for financial support.


\end{document}